\documentclass[a4paper,twoside]{revtex4}
\usepackage{epsfig}
\usepackage{hyperref}
\begin{document}
\title{Central peak in the pseudogap of high T$_c$ superconductors}
\author{D.~K.~Sunko}
\email{dks@phy.hr}
\author{S.~Bari\v si\'c}
\affiliation{Department of Physics, Faculty of Science, University of
Zagreb,\\ Bijeni\v cka cesta 32, HR-10000 Zagreb, Croatia.}
\begin{abstract}
We study the effect of antiferromagnetic (AF) correlations in the three-band
Emery model, with respect to the experimental situation in weakly underdoped
and optimally doped BSCCO. In the vicinity of the vH singularity of the
conduction band there appears a central peak in the middle of a pseudogap,
which is in an antiadiabatic regime, insensitive to the time scale of the
mechanism responsible for the pseudogap. We find a quantum low-temperature
regime corresponding to experiment, in which the pseudogap is created by
zero-point motion of the magnons, as opposed to the usual semiclassical
derivation, where it is due to a divergence of the magnon occupation number.
Detailed analysis of the spectral functions along the $(\pi,0)$--$(\pi,\pi)$
line show significant agreement with experiment, both qualitative and, in the
principal scales, quantitative. The observed slight approaching-then-receding
of both the wide and narrow peaks with respect to the Fermi energy is also
reproduced. We conclude that optimally doped BSCCO has a well-developed
pseudogap of the order of 1000~K. This is only masked by the narrow
antiadiabatic peak, which provides a small energy scale, unrelated to the AF
scale, and primarily controlled by the position of the chemical potential.
\end{abstract}

\maketitle


\section{Introduction}

The Fermi surface phenomena in the high-T$_c$ cuprates, and especially BSCCO,
have been extensively investigated, and a broad consensus has developed
concerning their main features. The Fermi surface is large and hole-like, with
a simple topology of a rounded square, or barrel, centered around the
M-point~\cite{Fretwell00}. Single-particle phenomenology is routinely invoked
on the ARPES spectra, thus the `self-energy' and `damping' are often extracted
from the main peak as if it were a coherent, weakly perturbed
quasiparticle~\cite{Valla99}. However, in the underdoped regime and below a
temperature scale T$^*$, the metallic state is surfeit with low-energy
correlations, about whose relevance for either the pseudogap, or the
superconducting mechanism itself, there is no general agreement at present.
Various experimental observations at low energy have been interpreted in terms
of stripes~\cite{Kivelson03}, paramagnons~\cite{Sadovskii01},
phonons~\cite{Lanzara01}, and superconducting fluctuations above
T$_c$~\cite{Ong04}. All these correlations are at present the object of
intense scrutiny, mainly with a view to ascertaining whether they enhance or
suppress superconductivity.

Theoretical understanding of the measured electronic spectral functions of
high-T$_c$ superconductors has received significant attention in the context
of these efforts~\cite{Shen95,Kim98,Schabel98,Fedorov99,Feng01,
Kaminski01,Mesot01,Matsui03,Campuzano04}. Physically, conduction occurs in the
copper oxide planes, so the most important electronic states are directly
accessible to surface probes such as ARPES. This naturally allows for a
concentration of theoretical effort, especially because the observed spectra
offer some outstanding puzzles of their own. Such a long-standing issue is the
appearance of the pseudogap~\cite{Randeria98}, observed near the vH points for
underdoped systems, and its connection with the AF gap at lower doping on the
one hand, and with the superconducting (SC) gap at lower temperature, on the
other. Experimentally, the pseudogap is clearly connected with a $(\pi,\pi)$
correlation~\cite{Campuzano04}, and the most natural candidate for its origin
are antiferromagnetic fluctuations above their transition
point~\cite{Sadovskii01}.

The present work attempts to connect several aspects of the low-energy
phenomenology of the cuprates in the hope of realistically constraining the
eventual theory of the optimally doped and weakly underdoped state. We adopt
an effective weak-coupling framework, and concentrate on aspects least
sensitive to model details. Our most important observation is that the
pseudogap does not really disappear at optimal doping, but is instead rather
inefficient at suppressing part of the spectral strength around the vH points.
The unsuppressed strength appears at the Fermi level as an `antiadiabatic'
central peak in the middle of a still fairly wide and deep pseudogap. It is
the latter `high-energy pseudogap' which indicates the underlying physical
scale, while the `leading edge' scale, associated with the central peak, turns
out to be incidental to the dynamics. It is primarily controlled by doping.
This interpretation does not even depend on the pseudogap being due to
antiferromagnetism as such, but only on the fact that the dominant perturbing
correlation occurs around the wavevector $\mathbf{Q}=(\pi,\pi)$. It is however
different than interpreting the high-energy `hump' in terms of bilayer
splitting~\cite{Feng01,Chuang01,Kordyuk02}.

We do not enter here the important question why the magnetic correlations
undergo an essential change at T$_c$. Our main aim is to show that when their
observed low-temperature behavior is introduced phenomenologically in the
calculation of the single-electron propagator, the resulting antiadiabatic
peak and pseudogap behave consistently with the main features of the
`peak-dip-hump' structure, found in experiments on superconducting optimally
doped BSCCO. In this way our calculation refers to the superconducting state.
We only omit the direct effect of superconductivity on the single electron
propagation, namely the appearance of a superconducting gap. This is justified
by the fact that the SC gap scale in ARPES is an order of magnitude below the
AF scale, manifested by the high-energy `hump.' In order to reproduce typical
normal-state ARPES profiles, which do not show a narrow low-energy peak, we
only need to overdamp the paramagnons. Our work provides a connection between
the observed simultaneous appearances of a magnetic resonance and of a narrow
low-energy peak in the ARPES profile, as the temperature drops below T$_c$.

Like some other authors~\cite{Vilk97,Friedel02,Eschrig03}, here we use
an effective weak-coupling (single-band) approach to describe the effect of
antiferromagnetic correlations on the single-electron propagation. Given that
$U_d$ is large in the high-T$_c$ superconductors, our starting point is the
strong coupling limit, and we use the present calculation to develop a
phenomenological framework in which the correct physical regime can be
identified for the effective weak-coupling approach. Section~\ref{sb} is thus
devoted to placing the present work in this wider theoretical context.
Section~\ref{reg} describes the model results. A comparison with experiment is
found in Section~\ref{exp}. Finally, Section~\ref{dis} is a recapitulation and
discussion.

\section{The electron self-energy and the central peak}\label{sb}

\subsection{Separation of charge and spin channels}

We enter a brief theoretical discussion now on the validity of the
weak-coupling single-band approach, with a large hole-like Fermi surface, when
$U_d$ is large. This is the essential input in our calculation, important for
its comparison with the $k$-dependencies measured by ARPES.

Recently a considerable improvement in understanding the band dispersion
measured by ARPES in the high-T$_c$ superconductors near optimal doping has
been achieved by considering the extended Emery model~\cite{Emery87} in the
limit of large interactions $U_d$ on the Cu-site. The original Emery model of
the CuO$_2$ plane is extended by taking into account the direct O-O hopping
$t'<0$ in addition to the original Cu-O hopping $t_0$ and the difference
$\Delta_{pd}$ of the O and Cu site energies~\cite{Mrkonjic03}.

In the limit of interest $|t'|>t_0^2/\Delta_{pd}$ this means that the `broad'
oxygen band is weakly hybridized with the Cu level. The Emery model then
resembles the Anderson lattice model which includes the accurate symmetry of
the electron (hole) propagation in the CuO$_2$ lattice, either along the O-O
axis ($t'$) or along the Cu-O axis ($t_0$). Notably, the limit $|t'|\gg
t_0^2/\Delta_{pd}$, although probably too extreme for physical purposes,
corresponds to the Falicov-Kimball model~\cite{Falicov69}, also sometimes
invoked in the context of high-T$_c$
superconductors~\cite{Gorkov89,Sunko96,Sunko00,Freericks01,Sunko05}.

The large $U_d$ limit of the Emery model~\cite{Emery87} extended by $t'$ was
treated in the homogenous mean field
approximation~\cite{Mrkonjic03} applied to the slave-boson
representation of the $U_d=\infty$ Emery model. The usual objection that the
mean-field slave boson (MFSB) approximation breaks the local gauge invariance
required by the slave-boson theory was met~\cite{Mrkonjic03-2} by emphasizing
that the static mean field merely represents the slow component of the slave
boson field. This latter, allowed by local gauge invariance, only appears as
static when particular physical properties are calculated, most notably the
physical electron band dispersion. Thus the physical dispersion can be
represented by the usual non-interacting three-band dispersion, but with
strongly renormalized tight-binding parameters $\Delta_{pf}$ and $t$ instead
of $\Delta_{pd}$ and $t_0$, while $t'$ remains unaffected by the copper
on-site repulsion. Most importantly, in this way the observed regime
$\Delta_{pf}\approx 4|t'|>t$ naturally replaces the regime
$\Delta_{pd}>t_0>t'$, inferred from chemical valency analysis and
high-energy spectroscopy data. The formula for the antibonding electron
band is then
\begin{equation}
\varepsilon(\mathbf{k})=\frac{1}{3}\sqrt{P}\left(
\cos\frac{\Psi}{3}+\sqrt{3}\sin\frac{\Psi}{3}\right),
\label{mfsb}
\end{equation}
where
\begin{eqnarray}
\Psi&=&\arccos\frac{Q}{P^{3/2}},\nonumber\\
P&=&12t^2f_1+48t'^2f_2+\Delta_{pf}^2,\nonumber\\
Q&=&144t'(3t^2+t'\Delta_{pf})f_2-\Delta_{pf} (18t^2f_1+ \Delta_{pf}^2),
\nonumber
\end{eqnarray}
with $f_1=\sin^2k_x/2+\sin^2k_y/2$ and $f_2=\sin^2k_x/2\sin^2k_y/2$. It is
obvious from Eq.~(\ref{mfsb}) that the effective near neighbor hoppings $t$
and $t'$ enter the dispersion $\varepsilon(\mathbf{k})$ non-linearly. This is
in contrast to those one-band approaches which include hoppings to
unphysically~\cite{Mrkonjic04-1} distant neighbors, but as independent
parameters. (Noteworthily, LDA calculations~\cite{Andersen95} show directly
that when such long-range hoppings are induced by reduction of multiband to
single-band models, the effective parameters depend non-linearly on the
near-neighbor ones, as in our case.) We use the single band (\ref{mfsb}) from
the three-band model with this distinction in mind.

Actually, after taking into account the fast harmonic slave-boson fluctuations
around the mean-field saddle point, the MFSB band (\ref{mfsb}) decomposes into
the narrow resonant band with dispersion $\varepsilon(\mathbf{k})$ and a
spectral density $A_\mathbf{k}$, which accomodates approximately $\delta$
holes (doping $\delta>1$) on the O-site and one hole localized in the
localized state on the Cu-site at the energy $\Delta_{pd}$, deep below
$\varepsilon(\mathbf{k})$~\cite{Mrkonjic03,Mrkonjic03-2}. The observed
structure of the resonant band in LSCO is well described~\cite{Mrkonjic03} by
the regime $|t'|>t_0^2/\Delta_{pd}$ and evolves with doping $\delta$ according
to MFSB predictions.

\begin{figure}[h]
\epsfig{file=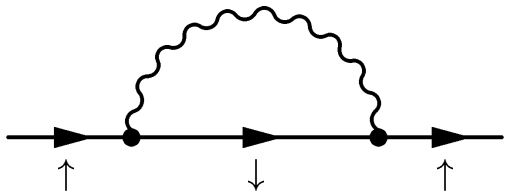,height=30mm}
\caption{Phenomenological one-magnon approximation.}
\label{figphen}
\end{figure}

Once these renormalizations are taken into account, the remaining low-energy
analysis concerns only the resonant band $\varepsilon(\mathbf{k})$ containing
the Fermi level. In the above slave-boson calculations the spins of the
localized holes on the Cu sites are taken as paramagnetic. This is justified
for large enough dopings $\delta$, when the band-width of the resonant band
exceeds the magnetic and/or superconducting energy scales, whereas for
$\delta\approx 0$ models of $t$-$J$ type may be more appropriate. Indeed, for
optimal dopings the bandwidth is of the order of 1~eV, whereas the magnetic
and/or superconducting effects occur on scales lower by at least an order of
magnitude. This approach is in principle well suited to take magnetic energies
into account as a perturbation of the main energy scales associated with the
resonant band. While the formation of the resonant band is associated in the
first place with the slow component in the motion of the slave boson, its fast
component is alone responsible for the weak magnetic couplings. Several
calculations of this type were carried out before for strongly interacting
electron systems. In particular, for the $t'=0$ Emery model with large $U_d$,
the residual effective couplings were derived explicitly~\cite{Tutis94}. The
extension of these ideas to finite $t'$, in particular to
$|t'|>t_0^2/\Delta_{pd}$, with a clear distinction between slow and fast
components of the slave-boson field is currently under way. These residual
couplings can be treated, for example, by the perturbational 2D parquet
theory. At $t'=0$ the latter was shown~\cite{Dzyaloshinskii88} to lead to
ladder-like results (`fast parquet') in most of the space of coupling
parameters, and to the marginal Fermi liquid only under very special
conditions. Here we consider this ladder-like regime appropriate for
calculating the electron self-energy $\Sigma$ from Fig.~\ref{figphen}, where
the wavy line represents the spin susceptibility $\chi$, and the triangular
vertex corrections are neglected.

To summarize, for large $U_d$ there is a natural separation of the slave-boson
fluctuations into fast and slow components. The latter appear static when
calculating the effective dispersion of the electrons, so in fact our
mean-field slave-boson renormalization of the electronic band parameters
corresponds to taking this slow component into account in the charge
channel~\cite{Mrkonjic03-2}. This explains why the Fermi surface is large and
hole-like. The fast component in the spin channel is the paramagnon
perturbation of the large, hole-like Fermi surface. In the following, we
concentrate entirely on the latter, neglecting triangular vertex corrections
in Fig.~\ref{figphen}.

\subsection{Electron spectral density}

Following the ideas expressed above, the fermion line in Fig.~\ref{figphen} is
taken to represent hole propagation in the absence of magnetic couplings.  In
this approach the electron Green's function appearing in Fig.~\ref{figphen} is
free. It has been observed~\cite{Vilk97-1,Yanase04} that corrections from
self-consistency tend to cancel with vertex corrections in the absence of
Migdal's theorem, so it appears generally more reliable to take neither into
account than only one. Thus the retarded Green`s function is just
\begin{equation}
G^{(0)}_R(\mathbf{k},\omega)=\frac{A_\mathbf{k}}{
\omega-\varepsilon(\mathbf{k})+\mu+i\eta},
\end{equation}
with $\varepsilon(\mathbf{k})$ from Eq.~(\ref{mfsb}), and $A_\mathbf{k}$ the
spectral density of the resonant band. Based on the above MFSB considerations,
we expect a significant $\mathbf{k}$-dependence in this quantity along the
Fermi surface, once the strong on-site repulsion is explicitly taken into
account. In the present work, we concentrate entirely on the vicinity of the
vH point, so $A_\mathbf{k}$ will eventually be absorbed into a coupling
constant.

The wavy line in Fig.~\ref{figphen} is taken to correspond to the simplest
form of the magnetic propagator,
\begin{equation}
\chi_R(\mathbf{Q+q},\omega)=\frac{\omega_0^2}{
(\omega+i\gamma)^2-\omega_D(\mathbf{q})^2},
\label{chi}
\end{equation}
where $\mathbf{Q}$ is close to the AF wave vector $(\pi,\pi)$, $\gamma$ is the
damping, and $\omega_D$ the dispersion
\begin{equation}
\omega_D(\mathbf{q})^2=\widetilde{\omega}^2+c^2|\mathbf{q}|^2.
\label{omegad}
\end{equation}
Here $\widetilde{\omega}$ is the band-edge, and $c$ the paramagnon velocity.
An upper cutoff $\omega_0$ to the magnons is also introduced, corresponding to
the extension $\omega_0/c$ of the magnon anomaly around $\mathbf{Q}$.

The static magnetic structure factor related to Eqs.~(\ref{chi})
and~(\ref{omegad}) is characterized by the value of $\mathbf{Q}$ and physical
correlation length $\xi=c/\widetilde{\omega}$. This structural factor is
measured directly by the elastic neutron
scattering~\cite{Rossat-Mignod91,Ishida96} or indirectly through the nuclear
spin relaxation rate $T_1^{-1}$. Both these types of experiment were recently
shown~\cite{Gorkov03} to be mutually consistent (in LSCO) when related by the
static limit of Eq.~(\ref{omegad}). For simplicity however, the incommensurate
effects (usually associated with `stripes') will not be included in the
present calculation, as they do not seem to be important for ARPES results.
The value of $\mathbf{Q}$ is therefore taken to be $(\pi,\pi)$ in
Eq.~(\ref{chi}), and $\xi$ is isotropic.

Turning further to the dynamical features of Eq.~(\ref{chi}), it should be
noted that the possibility of a central peak in the \emph{magnon} response,
below $\widetilde{\omega}$, appearing together with the strong dispersive
branches, is not included. We could include it fairly easily, following an
ansatz~\cite{Bjelis75} slightly different from  Eq.~(\ref{chi}). As shown
below, such slow correlations (usually associated with `dynamical stripes')
are not needed to reproduce the main features of the electronic spectral
structure observed by ARPES along the $(\pi,0)$--$(\pi,\pi)$ line. At fixed
$c$ and low temperature, the main parameter tuning the AF dynamics in
Eq.~(\ref{chi}) is the ratio of the damping $\gamma$ and the band-edge
$\widetilde{\omega}$. Magnetic fluctuations are strongly overdamped in the
normal state, but as soon as superconductivity sets in, a resonance peak
appears at 41~meV, around optimal doping~\cite{Rossat-Mignod91,Ishida96}.
Notably, Morr~\cite{Morr00} has obtained the magnetic resonance peak,
observed below T$_c$, from a mode with the dispersion~(\ref{omegad}) and
$\widetilde{\omega}\approx 20$~meV, simply by changing from overdamping to
underdamping.

The formal expression for the (retarded) fermion
self-energy~\cite{Abrikosov75} may then be rewritten as a sum of two
terms\footnote{If the lattice constant $a\equiv 1$, then the product
$g^2_{\mathbf{k,q}}A_\mathbf{k}$ is of dimension energy.}
\begin{eqnarray}
\Sigma_R(\mathbf{k},\omega)=-\frac{1}{2\pi^2}
\int g^2_{\mathbf{k,q}} d^2q\int_{-\infty}^\infty d\omega'&&\left[
\chi_R(\mathbf{Q+q},\omega-\omega')(1-f(\omega'))
\mathrm{Im}\,G^{(0)}_R(\mathbf{k-q-Q},\omega')
\right.
\nonumber\\&&
\left. +G^{(0)}_R(\mathbf{k-q-Q},\omega-\omega')
n(\omega')\mathrm{Im}\,\chi_R(\mathbf{Q+q},\omega')\right],
\label{sigma}
\end{eqnarray}
where $g_{\mathbf{k,q}}$ is the effective interaction vertex in
Fig.~\ref{figphen}, and $\mathbf{Q}=(\pi,\pi)$. The first term in
Eq.~(\ref{sigma}) is the boson propagator convoluted with the electron
response, the second, vice versa. In the high temperature limit the first is
negligible, because the Bose occupation term dominates the Fermi factor,
$n(\omega')\approx kT/\widetilde{\omega}\gg 1-f(\omega')$. In the
low-temperature limit $kT<\widetilde{\omega}$, which we consider here, both
terms may be equally important, with contributions coming from magnon
zero-point motion. The antiadiabatic central peak is due to the second (boson
response) term in both temperature limits, as is the lower side wing,
corresponding to occupied states. The effect of the first term in the
low-temperature regime is twofold: it provides the upper wing (unoccupied
states) of the pseudogap, and significantly affects the position and spectral
intensity of the peaks coming from the second term.

Each of the two terms can itself be expressed as the sum of two contributions,
a `dispersive' part from the propagator poles, and a `diffusive' part from the
poles in the occupation factors. The diffusive terms are proportional to the
damping in the respective response functions. They are not essential for the
physics discussed in the present work, although we include them in the
numerics, when we compare with experiment. The dispersive parts are
responsible for both the gap and the pseudogap, when it appears. As shown
below, the pseudogap $\Delta_{PG}$ can appear not only in the high-
($kT>\widetilde{\omega}$), but also in the low-temperature
($kT<\widetilde{\omega}$) regime, while the true gap is always in the
high-temperature limit, since in the present model it requires
$\widetilde{\omega}\to 0$ before $kT\to 0$. The temperature is taken to be
10~meV, lower than the other parameters in the problem.

As already emphasized in previous work~\cite{Bjelis75,Sunko93}, a special
physical regime applies in the vicinity of the vH point, where the electrons
themselves are slow, in fact static at the vH point itself. Then a frequency
`window' appears, of the order of the band-edge $\widetilde{\omega}$, in which
a weakly damped peak survives. This creates an `antiadiabatic' central peak in
the middle of the pseudogap, as long as the paramagnon band-edge is finite.
For a thorough example of the usual adiabatic regime from the side of broken
translational symmetry, see Ref.~\cite{Brazovskii76}, while
Ref.~\cite{Bjelis75} describes the antiadiabatic regime without translational
symmetry breaking. Both of these study charge density waves. In the high-T$_c$
context, Ref.~\cite{Sunko93} discusses the high-temperature antiadiabatic
case, while Ref.~\cite{Vilk97} is concerned with shadow-band signatures found
in the high-temperature overdamped regime ($kT>\gamma\gg\widetilde{\omega}$),
which we can also reproduce. The latter two references~\cite{Vilk97,Sunko93}
take the translational symmetry to be unbroken, like the present work.

\begin{figure}
\epsfig{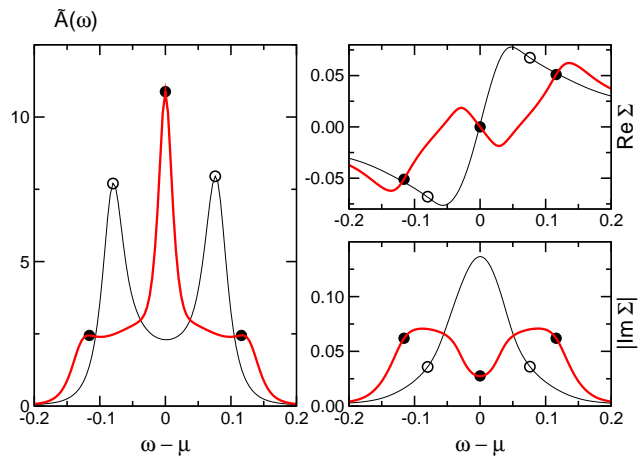}
\caption{Thick lines: antiadiabatic central peak in the renormalized electron
spectral density $\widetilde{A}(\omega)$ at the vH singularity
$\mathbf{k}=(\pi,0)$ with $\mu=\varepsilon_{vH}$, for a particularly simple
parametrization. Thin lines: $\mathbf{k}=(\pi/2,\pi/2)$. The circles have the
same abscissas in all three panels.}
\label{figpedagog}
\end{figure}
The appearance of the antiadiabatic central peak is first illustrated in the
renormalized spectral density $\widetilde{A}(\omega)$ of Fig.~\ref{figpedagog}
for a particularly simple parametrization. We put $t'=0$ and
$\mu=\varepsilon_{vH}$, so that the unperturbed system has a square nested
Fermi surface touching the vH singularity. The product
$g^2_{\mathbf{k,q}}A_\mathbf{k}$ is taken as constant, independent of the
position on the Fermi surface. The band-edge
$\widetilde{\omega}=0.03$~eV$>kT=0.01$~eV$\approx \gamma=0.015$~eV is set to
the low-temperature regime, as in the rest of the article, while the remaining
parameters are irrelevant for the discussion. The thin line in the left panel
gives the spectral strength at $k=(\pi/2,\pi/2)$, simply split into two peaks.
They have quasiparticle signatures, as shown in the right two panels:
$\mathrm{Re}\,\Sigma$ has a negative slope at the peak positions, and
$|\mathrm{Im}\,\Sigma|$ is small.

The thick lines show the situation at $k=(\pi,0)$, the vH point, for the same
parametrization. The maxima in the side wings, also denoted by full circles,
are evidently incoherent: the corresponding $\mathrm{Re}\,\Sigma$ has a
positive slope, and $|\mathrm{Im}\,\Sigma|$ is large. Clearly a central peak
has survived at the vH point, protected by the antiadiabatic mechanism. To see
this, note that when the boson response is peaked around $\mathbf{q}\approx 0$
in Eq.~(\ref{sigma}), the main contribution to $\Sigma$ at $\mathbf{k}\approx
(\pi,0)$, the vH point, comes from electrons at $\mathbf{k-Q}\approx
(0,-\pi)$, the other vH point, where they are slow.

Thus the central peak consists of vH electrons which do not scatter because
they barely move, so for them even the slowest available paramagnons are
averaged out; this is the antiadiabatic regime. It is clear that it violates
the Fermi liquid paradigm, since $|\mathrm{Im}\,\Sigma|\neq 0$ at the Fermi
level despite $\mathrm{Re}\,\Sigma=0$. This is because electrons interact with
dissipative bosons. When the boson damping $\gamma$ is zero,
$|\mathrm{Im}\,\Sigma(\omega)|\equiv 0$ for $|\omega|<\widetilde{\omega}$ at
the vH point itself; this was checked analytically~\cite{Sunko93}. It is
further important to note that the antiadiabatic peak does not necessarily
appear at the Fermi level, since it has its own $\mathbf{k}$-dispersion. In
the zone, $\mathrm{Re}\,\Sigma$ and $|\mathrm{Im}\,\Sigma|$ for the
antiadiabatic peak behave similarly as for a quasiparticle, including the
reduced but finite quasiparticle weight. We shall see that the peak can reveal
its antiadiabatic origin nevertheless, by disappearing with changing
$\mathbf{k}$ when the electrons involved acquire a significant velocity.

Our calculation can support various physically motivated notions of a
pseudogap simply by performing it in different regions of parameter space.
Because of this, the actual pseudogap obtained in a particular calculation
with realistic parameters usually behaves as a transitional form between
intuitive limiting cases. For example, one might say that a pseudogap exists
only if the central peak does not cross the Fermi energy. However, that
behavior continuously transforms into the usual quasiparticle one simply by
increasing the energy of the paramagnon band-edge, and in fact the central
peak can easily cross $E_f$ before the side wings have disappeared. Similarly,
the notion that $\mathrm{Re}\,\Sigma$ has to have a positive slope for a side
wing to be called incoherent is less restrictive than requiring
$|\mathrm{Im}\,\Sigma|$ to be maximal at the peak position. Thus the latter
notion of incoherence is found as a limiting case, while the former appears
across a wide range of parametrizations. Finally, one could define the
pseudogap by the spectral weight having a minimum at the Fermi energy, and a
maximum in $|\mathrm{Im}\,\Sigma|$ at the same position. The split
quasiparticle (thin line) in the left panel of Fig.~\ref{figpedagog} is a
limiting case for such a definition, where all the peaks appearing are
coherent. However, these peaks do not cross the Fermi energy. This corresponds
to the most common understanding of a pseudogap, where a coherent
quasiparticle splits in two because of strong scattering at $\omega=\mu$,
which is the precursor to the new zone boundary when $\widetilde{\omega}\to
0$ at fixed temperature.

Let us narrow the usage of the term `pseudogap' now, to the one found relevant
for the present work. We do not call the thin line in the left panel of
Fig.~\ref{figpedagog} a pseudogap, in spite of the valley between the two
peaks, corresponding to a large $|\mathrm{Im}\,\Sigma|$. However, the peaks
themselves are `coherent,' as discussed above. Here we reserve the term
`pseudogap' for manifestly incoherent side wings, like the side wings of the
thick line, irrespective of the nature or presence of a central peak in the
middle. Notably, other parametrizations can give three coherent peaks at the
vH point, the two side ones like at the nodal point, and an antiadiabatic one
in the middle; in that case there is no pseudogap at all, in the language of
this article.

\section{Model regimes}\label{reg}

\begin{figure}
\epsfig{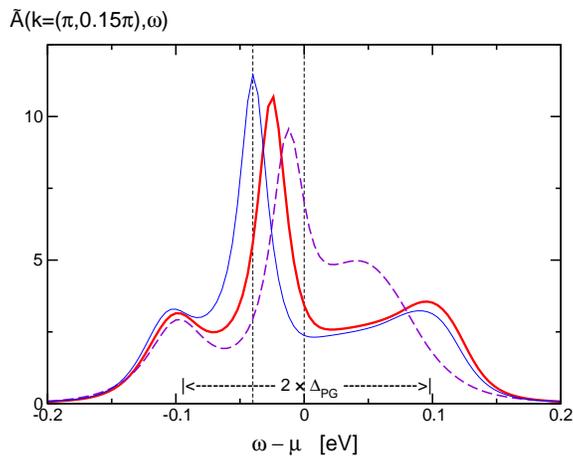}
\caption{Pseudogap with central peak at $k=(\pi,0.15\pi)$. Thick line:
parametrization in the text. Thin line: $\mu=0.035$~eV. Dashed line:
$kT=0.1$~eV, $F=0.026$~eV. Vertical dotted lines, left to right: paramagnon
scale ($\widetilde{\omega}=40$~meV), and Fermi energy.}
\label{figedc}
\end{figure}
At fixed low temperature, the ratio of the paramagnon `band-edge'
$\widetilde{\omega}$ to the damping $\gamma$ becomes the principal physical
parameter of the self-energy~(\ref{sigma}). We shall show below that this
number is relevant to account for the principal features of the ARPES
measurements in BSCCO and YBCO~\cite{Lu01} along the $(\pi,0)$--$(\pi,\pi)$
line, which we shall call X--M, in accord with the crystallographic notation
for YBCO. We take $\widetilde{\omega}$ to be 40~meV, in accord with
experiment, which puts our calculation in the low-temperature limit,
$kT<\widetilde{\omega}$. This means that the electrons are only perturbed by
paramagnon zero-point motion. We argue below it is this `quantum' pseudogap
which is actually observed in optimally doped BSCCO. The use of the term
pseudogap to refer to the destruction of the Fermi liquid behavior by magnetic
quantum fluctuations is already well established in studies of the
antiferromagnetic quantum critical point~\cite{Coleman99}.

The parameters in expression (\ref{sigma}) are treated semiphenomenologically,
\emph{i.e.} we shall use them primarily to adjust the experimentally observed
outcomes, but with regard to physically reasonable values. Let us give a
standard parametrization now, used throughout the article. The renormalized
parameters of the $U_d=\infty$ Emery model in the hole picture are:
copper-oxygen hopping $t=0.3$~eV, oxygen-oxygen hopping $t'=-1$~eV, and
effective copper-oxygen energy splitting $\Delta_{pf}=3.6$~eV. The Fermi
energy is in the electron antibonding band (\ref{mfsb}) of the three-band
model~\cite{Mrkonjic03}. The temperature is $kT=0.01$~eV, as already
mentioned.

The paramagnon parameters are the band-edge $\widetilde{\omega}=0.04$~eV,
damping $\gamma=0.015$~eV, cutoff $\omega_0=0.15$~eV, and correlation length
$\xi=c/\widetilde{\omega}\sim$3 lattice spacings. The coupling constant is
$g^2_{\mathbf{k,q}}A_\mathbf{k}\equiv F=0.077$~eV. Its wave-vector dependence
is neglected, because we concentrate on $\mathbf{k}\approx (\pi,0)$ and
$\mathbf{q}\approx 0$. To get a feeling for it, we note that the total range
of $\Sigma$ around the Fermi surface is $\sim$0.1~eV for this parametrization,
which is roughly an order below the width of the non-interacting antibonding
band~(\ref{mfsb}). Hence $100\times F/1$~eV may conveniently be imagined as
`percent' of the non-interacting electronic scale. The chemical potential is
$\mu=0.025$~eV from the vH point. (Larger $\mu$ means less holes.) Individual
parameter values are quoted elsewhere in the paper only to denote deviations
from the set given here.

The generic form of the pseudogapped spectral function with a central peak is
shown in Fig.~\ref{figedc}. As long as we are in the vicinity of the vH point,
some of the spectral strength of the slow electrons survives in the middle of
the pseudogap, itself of width $2\Delta_{PG}$, near the Fermi energy. The
persistence of a central peak in the single-loop approximation was noted
earlier~\cite{Bjelis75,Sunko93}. There appears an intrinsic `leading edge'
scale, the small distance from the central peak to the Fermi energy. When the
chemical potential is shifted toward underdoping, this distance increases. The
high background observed in ARPES does not appear here. It was obtained in
both ARPES~\cite{Melo90} and Raman~\cite{Niksic95} contexts by taking into
account the high-frequency slave-boson fluctuations in the charge channel,
which we do not consider here. (In the latter case~\cite{Niksic95} this was
done in the so-called non-crossing approximation, a somewhat different
starting point from the mean-field slave-boson one, on which this article is
based.) The upper and lower wings at $\Delta_{PG}>\widetilde{\omega}$ may be
understood in the semiclassical language as a consequence of the electronic
scattering on the nearly static, but still not completely ordered AF-like
potential induced by the paramagnons. This interpretation implies essentially
incoherent side wings, with a large $\mathrm{Im}\,\Sigma$ and
$\mathrm{Re}\,\Sigma$ with a positive slope. (In fact a different structure
can also appear, with coherent side peaks, as mentioned in the discussion of
Fig.~\ref{figpedagog}.) In the parametrizations used here to compare with
experiment, the side wings are in fact incoherent, while the physical regime
is at low temperature.

The pseudogap in the high-temperature limit looks quite similar, as shown by
the broken line in Fig.~\ref{figedc}. In particular the energy scale of the
side wings is easily adjusted to be the same. The qualitative behavior is
however different, and since the distinction is important for the
phenomenology, we discuss it now.

\begin{figure}
\epsfig{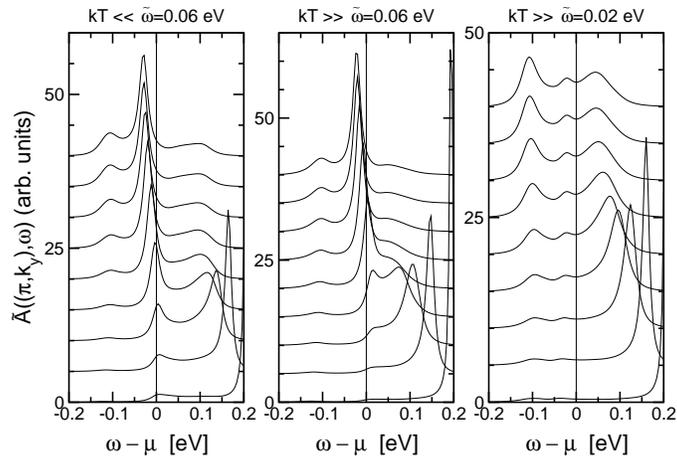}
\caption{Spectral functions along the X--M line, from the vH point (top
curves) to $k_x=\pi/a$, $k_y=0.4\pi/a$ (bottom). Left panel:
$\widetilde{\omega}=0.06$~eV. Middle: high-temperature limit ($kT=0.1$~eV,
$F=0.026$~eV) with $\widetilde{\omega}=0.06$~eV. Right: same high temperature,
$\widetilde{\omega}=0.02$~eV. (For high temperature, $F$ must be made smaller,
to compensate for the Bose occupation number giving an overall factor of
$kT$.)}
\label{figcomp}
\end{figure}
In Fig.~\ref{figcomp}, we show the calculated low-temperature (left panel) and
high-temperature (middle and right panels) peaks dispersing along the X--M
line. We first take the band-edge rather high, $\widetilde{\omega}=0.06$~eV,
to emphasize the antiadiabatic peak, which picks up most of the spectral
strength. Then the only difference between the left and middle panels is the
temperature, and we note that the antiadiabatic peak is much less dispersive
in the low-temperature case, and loses strength before crossing the Fermi
level. The low-temperature regime is influenced by both terms in
Eq.~(\ref{sigma}) equally, with a competition between the Bose and Fermi
contributions. In the high-temperature limit the second (boson response) term
takes over, with a different dynamics, due to the fact that when
$kT\gg\widetilde{\omega}>\gamma$, $n(\widetilde{\omega})$ in Eq.~(\ref{sigma})
becomes $\approx kT/\widetilde{\omega}$, introducing an additional dispersive
factor in the denominator. Note, however, that the intensities are given
without the Fermi occupation factor --- we show the renormalized spectral
density $\widetilde{A}(\omega)$, not $f(\omega-\mu)\widetilde{A}(\omega)$.
This means in particular that the loss of intensity in the left panel is not
due to the Fermi surface crossing, in fact we shall see (Fig.~\ref{figaad} in
the experimental section) that it occurs just as well when the antiadiabatic
peak stays away from the Fermi surface.

It is possible to keep the antiadiabatic peak below the Fermi energy in the
high-temperature regime as well, by lowering the band-edge
$\widetilde{\omega}$. This is shown in the right panel. Notice that lowering
the band-edge in the high temperature regime goes toward the opening of a true
gap, so the antiadiabatic signal is much smaller, relative to the side wings.
Otherwise, the dispersion is qualitatively similar to the left panel,
especially so when we realize that lowering the band-edge flattens the
dispersion in the low-temperature $kT<\widetilde{\omega}$ case as well (as
visible in Fig.~\ref{figaad}), similarly pushing the signal away from the
Fermi energy. The important qualitative difference in the behavior of spectral
strengths is however the following: in the left panel of Fig.~\ref{figcomp},
we notice that the side signal disappears before the antiadiabatic one; in the
right panel, they disappear together. In fact, the generic behavior in the
high-temperature $kT>\widetilde{\omega}$ limit is rather that the
antiadiabatic peak disappears sooner. We shall see in the next section that
experimental evidence in the superconducting state exhibits the
low-temperature behavior, providing one piece of evidence that the measured
response is in the low-temperature quantum regime.

\begin{figure}
\epsfig{file=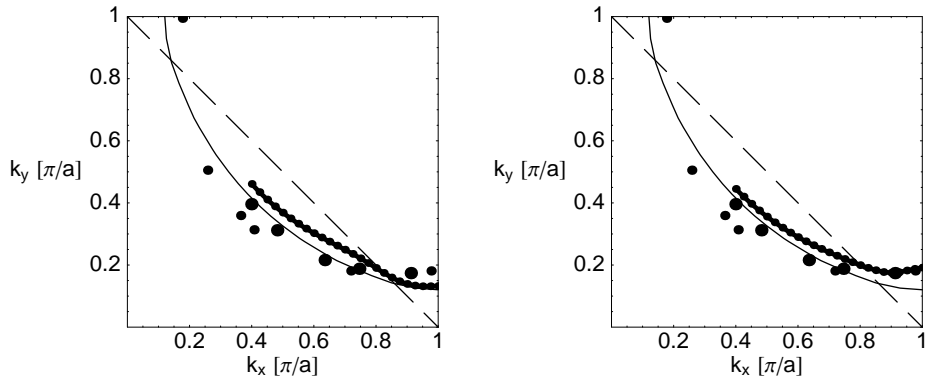,height=60mm}
\caption{Thin lines: zeroth-order  $U_d=\infty$ Fermi surfaces. Points
connected with thick lines: Fermi surfaces from maxima in momentum
distribution curves at $\omega=\mu$. Dashed lines: zone diagonals. Left:
high-temperature pseudogap, $kT=0.1$~eV and $F=0.026$~eV. Right:
low-temperature pseudogap. In this figure, $\mu=0.015$~eV throughout. Points:
experimental Fermi crossings, large: Ref.~\cite{Valla00}, small:
Ref.~\cite{Norman95}. (The value of $F$ in the high-temperature case is
adjusted to give a practically identical EDC profile at the vH point as in the
low-temperature case, in particular the same pseudogap scale.)}
\label{figfermis}
\end{figure}
The second piece of evidence is the effect of the magnon perturbation on the
Fermi surface. In the absence of a quasiparticle crossing, the experimental
community has developed various alternative criteria to define the Fermi
surface, one of which is the position of the maximum in the
momentum-distribution curve at fixed $\omega=\mu$~\cite{Kaminski01}. We adopt
that criterion in Fig.~\ref{figfermis}, which shows that the effect of magnon
perturbation on the zeroth-order Fermi surface is qualitatively different for
the high- and low-temperature pseudogaps. In the high-temperature
(semiclassical) regime $kT>\widetilde{\omega}$ (left panel), the tendency is
to change the shape of the Fermi surface so as to follow the zone diagonal in
the vicinity of the `hot spots'~\cite{Coleman99}, \emph{i.e.} the points of
intersection of the Fermi surface with the diagonal. Geometrically, this means
that the angle of the Fermi surface with the zone diagonal decreases. In the
low-temperature (quantum) regime $kT<\widetilde{\omega}$, shown in the right
panel, the result is precisely the opposite, the intersection angle increases,
and there is even a tendency to turn the effective Fermi surface upwards,
resulting in a `flared' shape. While our parametrization was chosen for a best
fit to energy distribution curves (see below), and we do not expect detailed
agreement with the Fermi surface shape, this qualitative difference is of
foremost physical importance. It shows, in effect, that `hot spot' scenarios,
for example Ref.~\cite{Yanase99}, correspond to the high-temperature regime of
Eq.~(\ref{sigma}). They depend on the similarity between `strong' and
`singular' scattering at the nesting wave vector, but this similarity is
qualitatively correct only when $kT>\widetilde{\omega}$, which is not the
observed case. The tendency of the Fermi surface to follow the zone diagonal
for $kT>\widetilde{\omega}$ is of course a precursor to the diagonal becoming
the new zone boundary, when the paramagnons condense. As already stressed
above, this can only happen in the present model when
$\widetilde{\omega}/kT\to 0$. While the upturn of the Fermi surface has never
been clearly observed --- there is only one experiment~\cite{Norman95}, the
small points Fig.~\ref{figfermis}, which seems to show such a tendency --- the
bending to follow the zone diagonal can be excluded with certainty. The Fermi
surface of optimally doped BSCCO in the vicinity of the vH points is at least
a straight line parallel to the $\Gamma$--X line. The marked difference
between the two final shapes (thick lines) in Fig.~\ref{figfermis} means that
in the present model one must choose the right-hand panel to reproduce the
experimentally observed shape. Thus, both the evolution of ARPES spectra in
the Brillouin zone and correction to the zeroth order $U=\infty$ shape of the
Fermi surface in BSCCO seem to point in the same direction, that the observed
pseudogap is due to zero-point motion of the magnons.

This discussion can be followed in the form of Eq.~\ref{sigma}. In the
semiclassical regime $kT>\widetilde{\omega}$, the first term is negligible
with respect to the divergence of the boson occupation number in the second
term, which is a precursor to the true gap. In the low-temperature regime, the
two terms are of the same order. They contribute even at zero temperature,
because the paramagnon zero-point motion can excite electrons within
$\widetilde{\omega}/2\approx 20$~meV of the Fermi energy, and since the vH
singularity is roughly within this range, that means a lot of them. Thus we
can violate the conventional Fermi liquid picture simply by putting in
dissipative paramagnons and letting $kT/\widetilde{\omega}\to 0$, keeping
$\widetilde{\omega}$ at its observed value, as discussed on the simple example
of Fig.~\ref{figpedagog}. It would of course be interesting to understand why
the paramagnon resonance should appear when the system goes superconducting.
We hope to shed more light on this question when we consider the on-site
repulsion explicitly, as mentioned in Sect.~\ref{sb}.

\section{Comparison with experimental spectra}\label{exp}

\begin{figure}
\epsfig{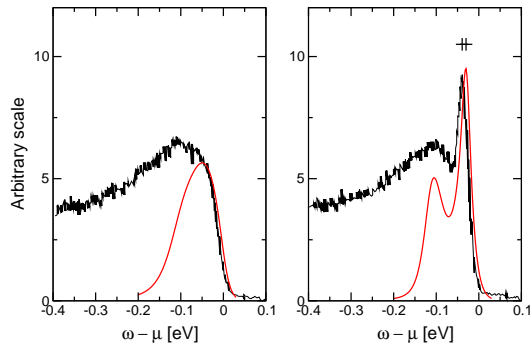}
\caption{Comparison with experiment on BSCCO (jagged line~\cite{Fedorov99}),
left: T=103~K, right: T=46~K. Calculation (smooth curves), left: overdamped
magnons ($\gamma=0.06$~eV), right: underdamped magnons (parameters in the
text). The scale $kT_c\sim 100$~K is approximately given by the double thin
cross, which marks the difference in position between the measured and fitted
narrow peaks.}
\label{figpi0}
\end{figure}
In Fig.~\ref{figpi0}, we show the comparison with experiment, used to
establish the parametrization. The jagged lines are measured ARPES
intensities~\cite{Fedorov99} of optimally doped BSCCO, integrated along the
$(0,0)$--$(\pi,0)$ line, in the normal (T=103~K, left) and superconducting
state (T=46~K, right). The smooth lines are both calculated at the $(\pi,0)$
point with the parameters in the text, the only difference being the
paramagnon damping: left, overdamped ($\gamma=0.06$~eV), right, underdamped
($\gamma=0.015$~eV). The difference, marked by crosses, in the positions of
the leading peaks in the right panel is $\sim$100~K, the superconducting
scale. The vertical scales of the two experimental and the two theoretical
curves are the same, while the relative scale of theory vs. experiment is
arbitrary. The single (overdamped) peak in the left panel and the side peak in
the right panel correspond to a positive slope of $\mathrm{Re}\,\Sigma$ and
large $\mathrm{Im}\,\Sigma$, \emph{i.e.} are both incoherent. Only the narrow
(leading-edge) peak in the right panel is coherent, with negative slope of
$\mathrm{Re}\,\Sigma$ and small $\mathrm{Im}\,\Sigma$.

\begin{figure}
\epsfig{file=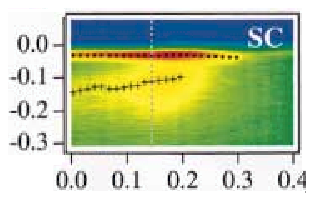,height=35mm}\hskip 1ex
\epsfig{file=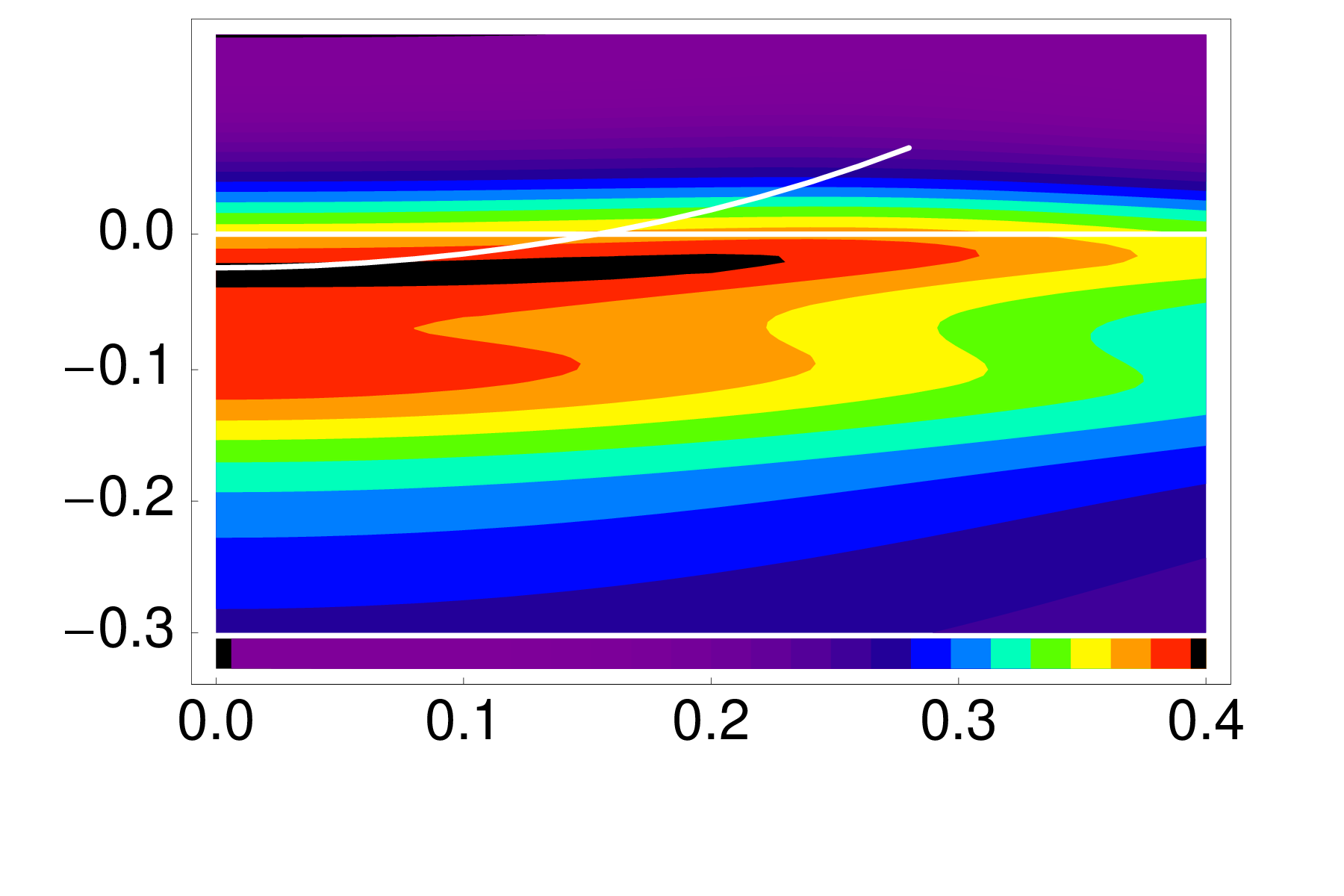,height=35mm}
\caption{Experimental log-intensity distribution in optimally doped
BSCCO~\cite{Kaminski01} (left), compared with the present calculation (right).
The vertical scale is in eV from the Fermi energy, the horizontal one in
$\pi/a$ along the X--M line. White lines: horizontal, Fermi energy; curved,
unperturbed $U_d=\infty$ dispersion. The intensities are multiplied by a Fermi
factor at 100~K.}
\label{figexp}
\end{figure}
Having fixed all parameters at the $(\pi,0)$ point in Fig.~\ref{figpi0}, we
now compare model predictions of evolution in the BZ with a different set of
experimental data~\cite{Kaminski01}, first in Fig.~\ref{figexp}. The strong
non-dispersive structure observed at the Fermi level corresponds to the
antiadiabatic peak in the calculation. The lower wing of the pseudogap is
shifted away from the position of the original three-band dispersion,
reproducing the experimental high-energy `hump' scale of $\sim$100~meV. The
ridge at $\sim$100~meV binding roughly follows the original dispersion, broken
and shifted by the paramagnon interaction. The antiadiabatic peak is
interpolated between it and the upper  wing (not visible because of the Fermi
factor). Notably, the same behavior has been reported along the X--M line in
YBCO~\cite{Lu01}.

\begin{figure}
\epsfig{file=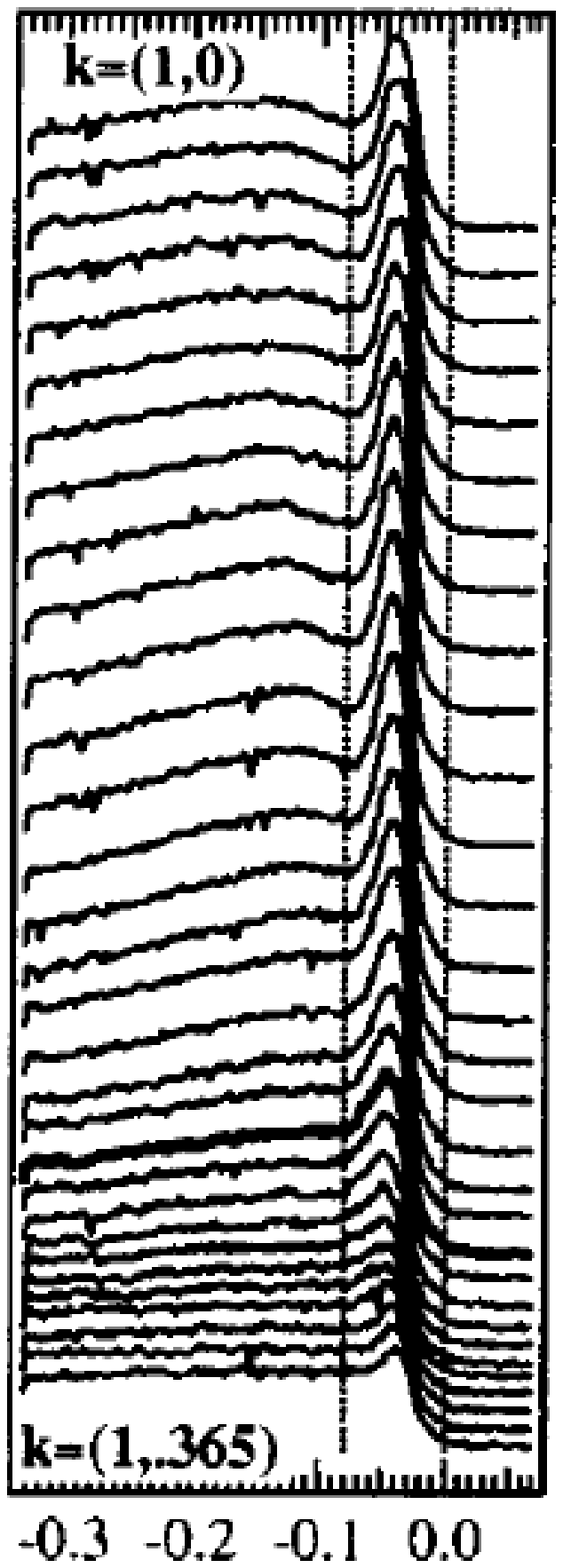,height=60mm}
\epsfig{file=figaad,height=60mm}
\epsfig{file=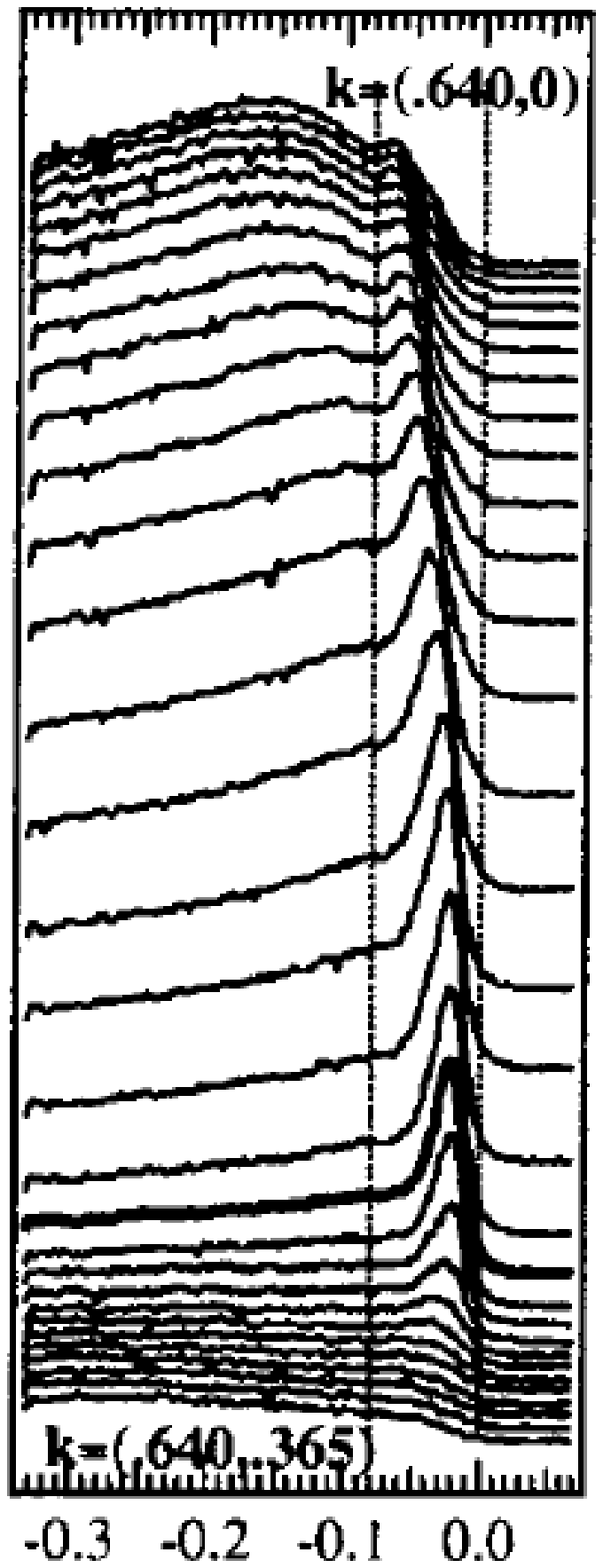,height=60mm}
\caption{Left to right: a) experimental EDC`s along the X--M line, from
$k_y=0$ (top) to $k_y=0.365\pi$ (bottom), multiplied by a Fermi function at
100~K and offset for clarity~\cite{Kaminski01}. b) Calculated intensities for
the same situation. c) the same for the line parallel to X--M at $k_x=0.9\pi$.
d) experiment for the line parallel to X--M at $k_x=0.64\pi$. EDC`s
corresponding to the Fermi surface crossing from the maximum in the momentum
distribution curve are given by a thicker line. The measurement and parameters
are the same as in Fig.~\ref{figexp}.}
\label{figaad}
\end{figure}
In Fig.~\ref{figaad}, we show the detailed energy distribution curves (EDC`s)
corresponding to Fig.~\ref{figexp}. All the qualitative experimental features
are correctly reproduced: both the major and minor energy scales, and the
downturn (in energy) of the antiadiabatic peak as one moves further away from
the Fermi crossing. Such an approaching-then-receding of the narrow peak with
respect to the Fermi level has been noticed in experiment~\cite{Kaminski01},
and becomes more pronounced with underdoping.

The reference position of the central peak at $(\pi,0)$ is at the Fermi
energy, as observed in our simple example in Fig.~\ref{figpedagog}. As soon as
the chemical potential is shifted, or one looks at other points in the BZ, the
peak moves away from the Fermi energy, producing a `leading-edge' energy scale
of the order of the chemical potential, affected of course by its own
dispersion. It is unrelated either to the primary AF scale, which determines
the width of the pseudogap by the `high-energy' side lobes, or the
superconducting scale, which does not appear in the calculation at all. Of
course, as the coupling constant decreases, or band-edge $\widetilde{\omega}$
increases, the peak begins to turn back into an ordinary (weak-coupling)
quasiparticle.

In the right two panels of Fig.~\ref{figaad}, we show what happens as one
moves towards the $\Gamma$ point in cuts parallel to the X--M line. We note a
significant redistribution of spectral strength, such that the side peak is
much stronger at $k_y=0$, the $\Gamma$--X line itself, but quickly loses
strength as one moves perpendicularly away from it in the $k_y$ direction,
parallel to the X--M line. Finally at $k_y=0.35\pi/a$, only the antiadiabatic
peak survives. Experimentally, much the same behavior has been observed, with
the proviso that it seems to evolve more slowly in the $\Gamma$ direction, so
the qualitative features we calculate around $k_x=0.9\pi$ here are observed
around $k_x=0.64\pi$ in experiment~\cite{Kaminski01}. Again, the wide peak has
its own approaching-then-receding sequence, similar to the one observed both
in optimally doped~\cite{Kaminski01} and underdoped
samples~\cite{Campuzano99}. The fact that the calculated qualitative features
continue to match closely the experimental situation as one moves away from
the X--M line into the zone interior, while the quantitative evolution
proceeds at a different pace, is possibly due to neglecting the
$\mathbf{k}$-dependence of the product $g^2_{\mathbf{k,q}}A_\mathbf{k}$ in the
calculation.

The principal outcome of the comparison with experiment is that the
experimental situation~\cite{Kaminski01} in the superconducting state
corresponds to the model regime of low temperature and low damping
$kT\approx\gamma<\widetilde{\omega}$. This allows us to claim that the
pseudogapped regime in fact extends to optimal doping. Furthermore, the fact
that our underdamped curves are calculated in the normal state means that the
main qualitative effect of superconductivity on the ARPES signal is due to the
reduction of magnon damping in the superconducting state.

\section{Summary}\label{dis}

The present work associates the observed spectra around the vH point with the
concept of an antiadiabatic central peak. It appears in the middle of a
pseudogap, representing that part of the spectral strength which is not
suppressed by the usual adiabatic mechanism of the opening of a pseudogap. The
underlying main AF scale $\Delta_{PG}$ is allways that of the side wings in
Fig.~\ref{figedc}, which is observed in ARPES as a `high-energy' feature, or
`hump.' In this way we are able to claim that the leading-edge scale,
connected with the narrow peak, is not due to any independent physical
phenomenon. The position of the antiadiabatic peak is sensitive to the
chemical potential, which naturally accounts in our scheme for the increase of
the `leading edge' pseudogap with underdoping. We can easily recover the
conventional Fermi liquid by raising the paramagnon band-edge or lowering the
coupling constant. At some point the `hump' scale goes to zero, and one
recovers the usual weakly perturbed quasiparticle. This is consistent with the
observation that the $\Delta_{PG}\sim$~T* scale disappears at overdoping,
rather than merging with the superconducting scale~\cite{Loram00,Ozyuzer03}.

The pseudogap with an antiadiabatic peak was found in this work to have two
physical regimes, low- and high-temperature, relative to the paramagnon
band-edge. The regime of low temperature corresponds to experiment in
optimally doped BSCCO. In the model, it gives rise to a dispersion for the
antiadiabatic peak which is qualitatively different from the bare $U_d=\infty$
one, amounting to a non-dispersive `feature' at a few tens of meV binding
energy. Thus the observed leading-edge scale need not be entirely due to the
superconducting gap. The low-temperature regime $kT<\widetilde{\omega}$ is
pseudogapped, because a true gap appears when $\widetilde{\omega}\to 0$ before
$kT\to 0$, \emph{i.e.} it is always in the high-temperature regime. As long as
$\widetilde{\omega}$ is held fixed, a pseudogap-like situation will occur for
$kT<\widetilde{\omega}$, without developing into a true gap even for the
lowest temperatures. However, it may be a true pseudogap, in the sense that
the side wings are incoherent, and this is the case in the parametrization
used here to compare with experiment. Lowering the band-edge in the
low-temperature regime makes the central peak disappear just like in the
high-temperature case, thus naturally accounting for the underdoped situation.

We took much trouble with Figs.~\ref{figcomp} and~\ref{figfermis} to choose
the low-temperature regime, although the magnon mode at 41~meV is obviously
much higher than the temperature. The reason is that our calculation is so
simple and generic that it fairly represents the perturbation by any bosonic
mode which does not have a slow component. There are a number of observed
low-energy fluctuations, such as stripes, which we do not take into account
here. At present, we cannot completely exclude a possible role of slow (spin)
fluctuations in the electron response measured by ARPES. These can be modelled
to some extent by a `central peak' in the boson response, distinct from the
dispersive branches studied here, and easily introduced in our calculation by
a slightly more general parametrization~\cite{Bjelis75} of
$\mathrm{Im}\,\chi$. However, since the observed spectral strength
redistribution and Fermi surface shape both correspond to our low-temperature
regime, we can ascribe the main features of the ARPES response to the
dispersive paramagnons. In this way the other low-energy phenomena are
relegated to a secodary role, possibly having to do with the shape and
spectral composition of the side wings. Even if the pseudogap were not due to
paramagnons, still the conclusion would remain that the characteristic energy
scale of the relevant boson is higher than the temperature, hence the
pseudogap need not of itself imply any additional ground-state phenomena. As
things stand, we see no reason to depart essentially from the natural
interpretation in terms of AF paramagnons.

The neglected quantum fluctuations in the charge channel are also expected to
affect the width and shape of the high-energy hump observed in ARPES, which
appears sharper and narrower in our calculation than in experiment. Apart from
that, the EDC`s obtained along the X--M line have a striking resemblance to
experiment in their main features. Both the high-energy scale of
$\sim$100~meV, and the leading-edge scale of 20--30~meV are correctly
reproduced. The main intensity pattern, where the peak loses strength further
away from the vH point, without ever crossing the Fermi energy, is reproduced
as well. The intensity shifts between the central peak and side wings are also
obtained. Finally, the observed slight variation in the narrow peak position,
which approaches the Fermi energy and then recedes from it, is also found in
the calculation. Thus we believe that we have understood the physical origin
of the narrow lowest-energy signal along the X--M line to be quite general:
the electrons giving rise to this signal are slower than the perturbing
paramagnons, and so escape the adiabatic suppression which opens the
pseudogap. We can easily recover the actual experimental situation for $T\gg
T_c$, simply by overdamping the paramagnons, which washes out the
antiadiabatic peak. This scenario is naturally consistent with the fact that a
precursor of the narrow peak is sometimes observed above T$_c$. Based on the
above discussion, we claim that the pseudogap in optimally doped BSCCO is in
fact fully developed, of the order of 1000~K, and is only masked by the
antiadiabatic peak. In this way we can view the superconducting correlations
as a third scale, an order of magnitude lower than the pseudogap `hump' scale,
and two orders of magnitude below the on-site repulsion, here taken into
account through the overall band renormalization. Their interplay with the
antiadiabatic leading-edge scale found here, which is of the same order of
$\sim$10~meV, should be of interest. It remains to be seen whether the
physical regime found here to be relevant for the low-energy cuprate
phenomenology can be consistently obtained from a microscopic strong-coupling
approach in the presence of an oxygen-oxygen overlap, as outlined in Section
\ref{sb}.

To conclude, we described the main low-energy signal along the X--M line in
optimally doped BSCCO to be due to an antiadiabatic central peak. The
pseudogap in the same material is an order of magnitude above the
superconducting scale, and persists below T$_c$. The small separation between
the antiadiabatic peak and the Fermi level appears naturally in the
calculation, even in the absence of explicit superconductivity. It is
primarily determined by the value of the chemical potential, in a way
consistent with its observed variation with doping.

\acknowledgments

Conversations with J.~Friedel and E.~Tuti\v s are gratefully acknowledged.
This work was supported by the Croatian Government under Project~$0119256$.


\end{document}